\documentclass[11pt,twoside]{article}

\usepackage{asp2004}
\usepackage{epsf}
\usepackage{psfig}
\usepackage{lscape}
\usepackage{graphicx}

\begin{document}
\title{Time resolved spectroscopy of the multiperiodic pulsating subdwarf B star PG1605+072}   
\author{A. Tillich$^1$, U. Heber$^1$ and S. J. O'Toole$^2$}
\affil{$^1$Dr.Remeis-Sternwarte Bamberg, Universit\"at Erlangen-N\"urnberg, Sternwartstr.7, D-96049 Bamberg, Germany\newline$^2$Anglo-Australian Observatory, P.O. Box 296 Epping, NSW 1710, Australia}

\begin{abstract}
We present results for the 2m spectroscopic part of the
MultiSite Spectroscopic Telescope campaign, which took place in May/June
2002. In order to perform an asteroseismological analysis on the multiperiodic
pulsating subdwarf B star PG 1605+072 we used over 150 hours of time
resolved spectroscopy in order to search for and analyse line profile variations by using
phase binning. We succeeded in finding variations in effective temperature and gravity
for four modes. A pilot analysis using the \textit{BRUCE} and \textit{KYLIE} programs and
assuming strong rotation and low inclination favours models with $l=1$ or $l=2$ with $m\leq0$.
\end{abstract}



\section{Introduction}
Subdwarf B stars are hot pre-white dwarfs ($T_{\mathrm{eff}}=22\mathrm{kK} - 40\mathrm{kK}$) in the core-helium burning phase of their evolution (Heber, 1986). A new class of non-radial pulsators (EC14026 or V361Hya stars) has been discovered (Kilkenny et al. 1997). The pulsations are driven by an opacity bump caused by iron ionisation (Charpinet et al., 1996).
PG1605 were discovered to be the largest amplitude pulsator amongst the V361Hya stars by Koen et al. (1998). More than 50 oscillation periods have been found in its light curve (Kilkenny, 1999). Kawaler (1999) suggested a rapid rotation ($v_{\mathrm{rot}}=130\mathrm{kms^{-1}}$) to explain the frequency splitting. Indeed a high projected rotational velocity $v\sin{i}=39\mathrm{kms^{-1}}$ has been measured (Heber et al., 1999). Time series spectroscopy revealed radial velocity variations (O'Toole et al., 2000) with more than 20 confirmed periods (O'Toole et al., 2005). In this paper we attempt to measure line profile variations and determine effective temperature and surface gravity variations in order to identify the individual pulsation modes.\\

\section{The MSST data and phasebinning}
In the 2m-part of the MSST campaign four observatories (Steward Observatory at Kitt Peak, ESO at La Silla, Siding Spring Observatory, NOT at La Palma) were involved. 9 nights of observation were followed by 12 more nights after a break of almost 3 weeks, resulting 10892 time resolved spectra (O'Toole et al., 2005). We treat the data sets of each telescope separately.\\
In the first step we reduced spectra using the IRAF package. For each spectrum the Doppler shift, was removed. After that a continuum is fitted to the spectra. The line profiles are supposed to change if the star is pulsating, depending on its phase. The main reasons for those changes are of course temperature and density perturbations in the stellar atmosphere, besides the radial velocity variations.
\setcounter{figure}{0}
To be able to detect these tiny variations, we determined the phase of the pulsation mode from the known periods for every single spectrum and coadded them accordingly. In this manner we got a complete pulsation cycle divided into twenty bins. Then we fitted LTE-model spectra to the binned spectra by performing a $\chi^2$-minimization using the \textit{FITPROF} program (Napiwotzki, 1999). So we were able to determine simultaneously the three atmospheric parameters ($T_{\mathrm{eff}}$, $\log{g}$ and $\log{He/H}$-ratio) for every bin. Their variations for the strongest mode f1 are shown in Fig. 1.
 \begin{figure}[h]
\begin{center}
 \includegraphics [scale=.4] {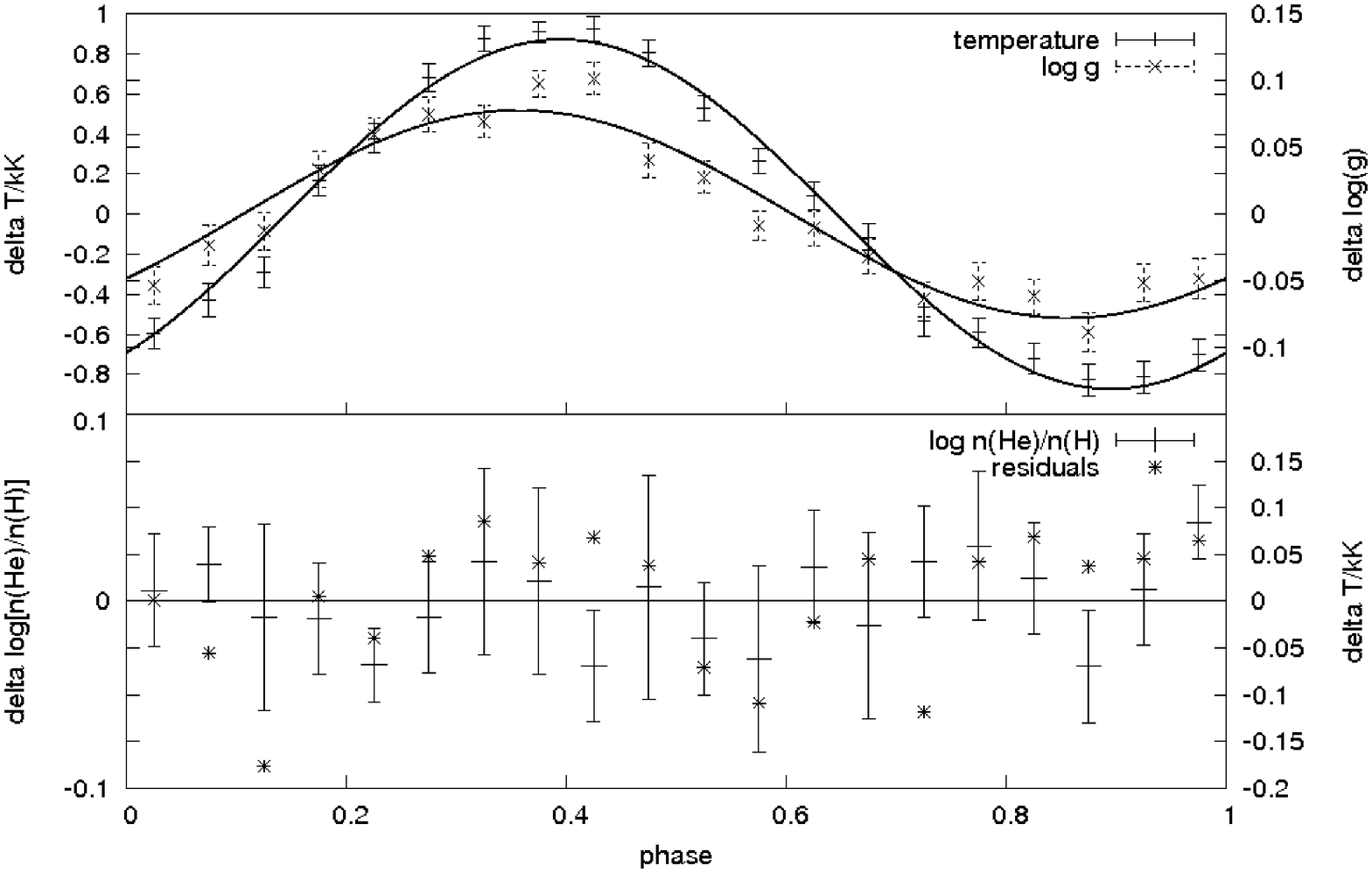}
\end{center}
\caption{\footnotesize{The variations of the atmospheric values determined from the Steward data with error bars and sine fit are shown. In the upper panel temperature and gravity with sine fits are shown. The lower panel shows the He/H abundance and the temperature residuals.}}
 \end{figure}
The temperature amplitude is $1750\mathrm{K}$, while the surface gravity variation is $0.16\mathrm{dex}$. The phase shift between the two fits is an indication for observing a nonradial mode. We also plotted the He/H abundance to make sure that it remains constant over a cycle.\\
Fig. 2a shows temperature variations of the four strongest modes.
\begin{figure}[h]
\includegraphics [scale=.26] {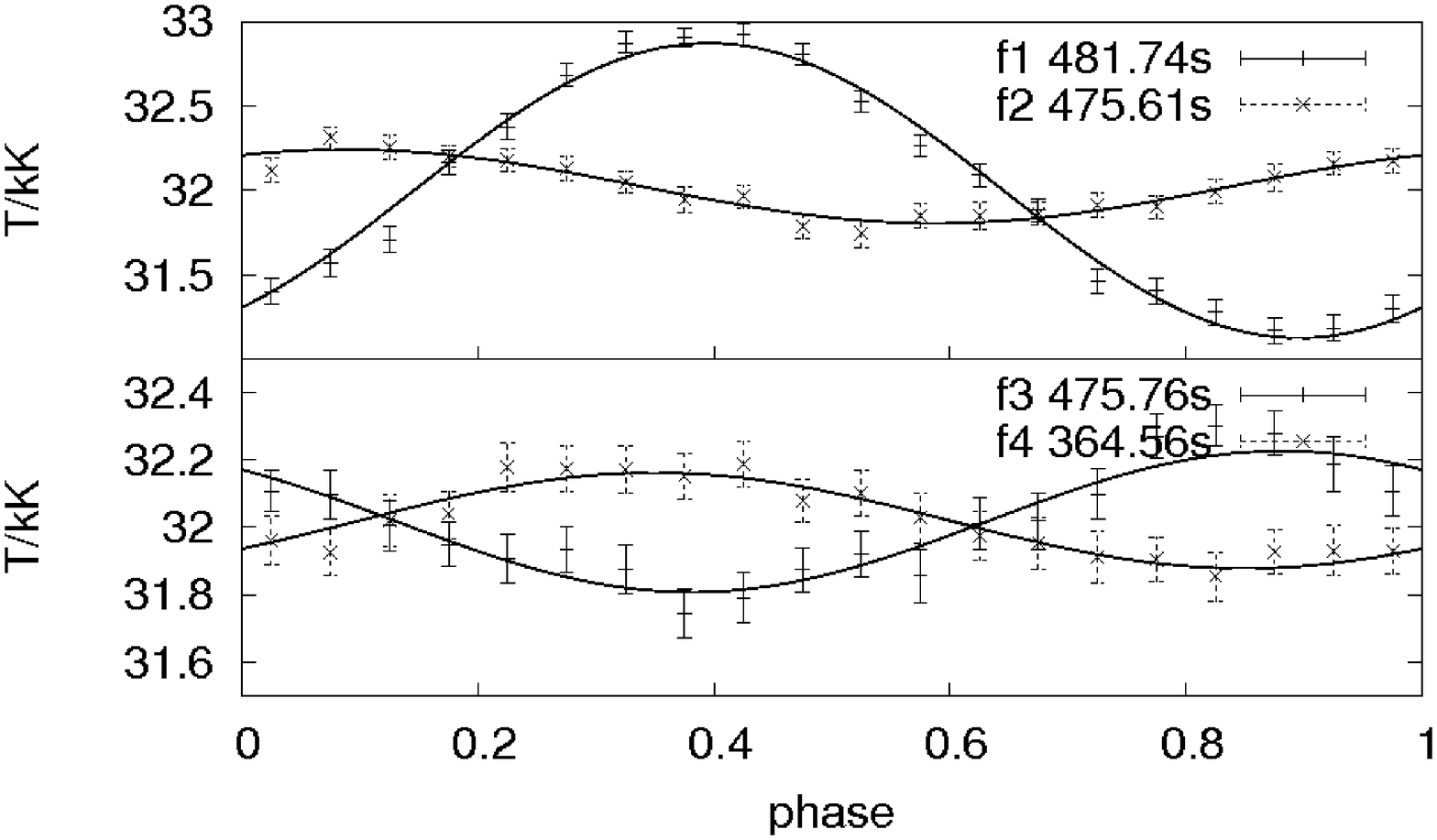} 
\includegraphics [scale=.26] {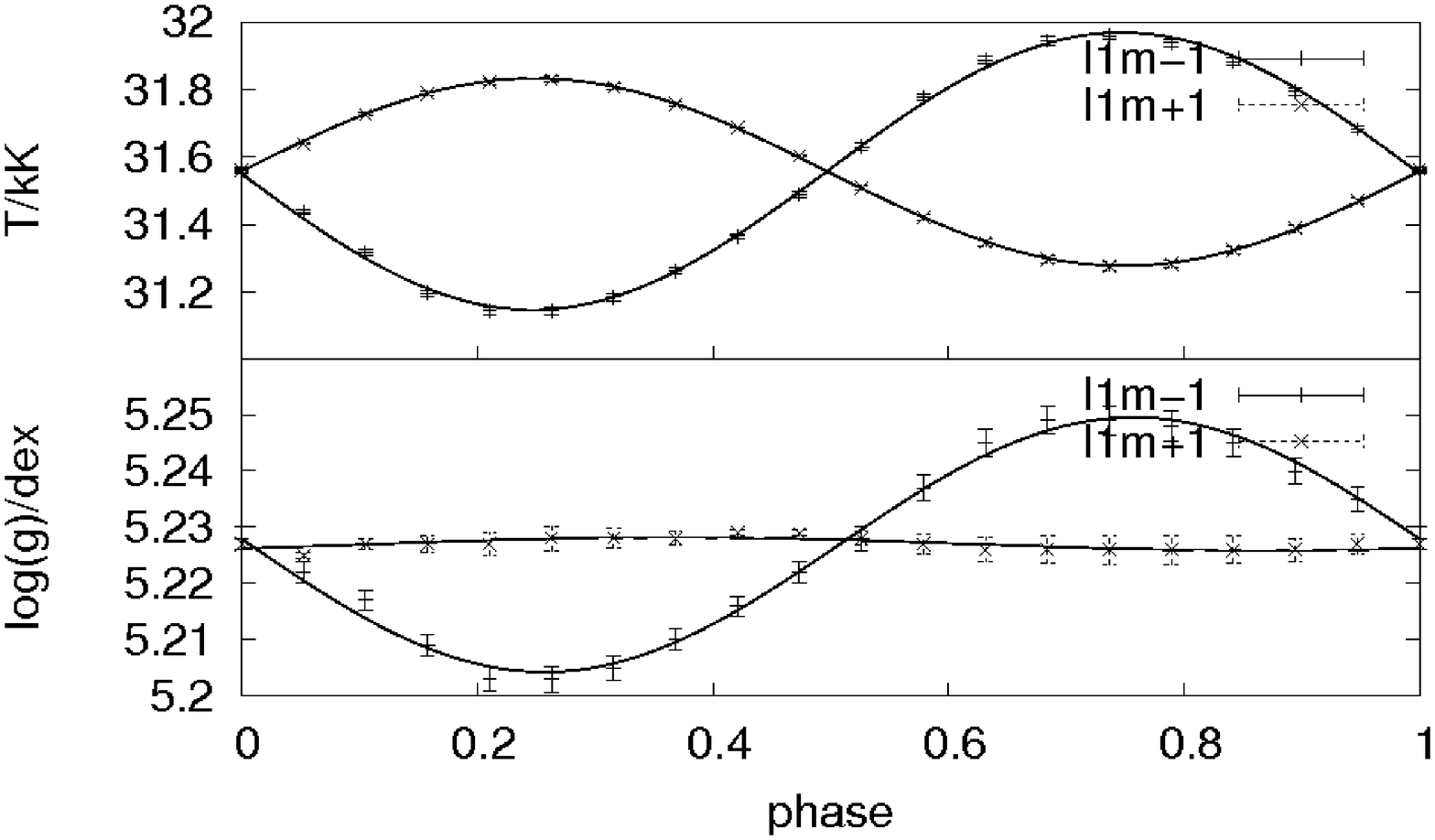}
\caption{\footnotesize{\textit{a (left panel)}: the temperature variations for the four strongest modes determined from the Steward data with error bars and sine fit are shown.\newline\textit{b (right panel)}: temperature and $\log{g}$ variations from synthetic spectra for the modes $l=1,m=\pm1$ with sine fits and error bars are shown.}}
\end{figure}
Fig. 2a and Table 1 show the amplitude of the pulsations is decreasing with decreasing radial velocity amplitude of the mode.
\begin{table}
\begin{center}
\caption{\footnotesize{Amplitudes of temperature, surface gravity and radial velocity variations for the four strongest modes.}}
\begin{tabular}{ccccc}
\tableline
\noalign{\smallskip}
\small Mode & \small $\pm\frac{\Delta T}{2}$ in K & \small $\pm\frac{\Delta\log{g}}{2}$ in dex & \small Period in s & \small $v_{\mathrm{rad}}^{\mathrm{max}}$ in $kms^{-1}$\\
\noalign{\smallskip}
\tableline
\noalign{\smallskip}
\small f1 & \small 873.7 & \small 0.078 & \small 481.74 & \small 15.4\\
\small f2 & \small 218.5 & \small 0.019 & \small 475.61 & \small 5.4 \\
\small f3 & \small 209.1 & \small 0.019 & \small 475.76 & \small 3.0 \\
\small f4 & \small 141.8 & \small 0.011 & \small 364.56 & \small 2.5 \\
\noalign{\smallskip}
\tableline
\end{tabular}
\end{center}
\end{table}

\section{Modelling of line profile variations and mode identification}
In order to identify the modes we had to model various pulsation modes. We used the \textit{BRUCE} and \textit{KYLIE} routines by R.Townsend (1997). \textit{BRUCE} calculates an equilibrium surface grid of the star's envelope for a given set of stellar parameters. For any predefined set of quantum numbers it calculates effective temperatures and gravities for $40000$ surface elements. KYLIE takes this grid and calculates synthetic spectra by interpolating in the same grid of model spectra used for the analysis of the observations. In the same way as with the observations, we used \textit{FITPROF} to determine the atmospheric parameters out of the synthetic spectra from
\textit{KYLIE} for every phase bin.\\
The required average atmospheric parameters ($T_{\mathrm{eff}}= 32\mathrm{kK}, \log{g} = 5.25\mathrm{dex}$) have already been determined by Heber et al. (1999). Following Kawaler (1999) we choose a high rotational velocity of $130\mathrm{kms^{-1}}$ and requiring a low inclination angle of $17^\mathrm{\circ}$ with a projected rotational velocity of $v\sin{i}=39\mathrm{kms^{-1}}$(Heber et al., 1999).\\
In the last step we tried to compare the calculated amplitudes and phase shifts between $T_{\mathrm{eff}}$ and $\log{g}$ with the measured ones. As there is almost no phase shift detected in the observations, modes with $m\leq0$ can be excluded.\\
The analysis showed that the radial mode $l=0,m=0$ has $\Delta T_{\mathrm{eff}}=3355.6\mathrm{K}$, which is much too high to be an observed mode, while the g-amplitude comes close to the observations. Further we discovered that for a fixed $l$ always the mode with $m=0$ is the strongest. For all the modes with $m\leq0$ the temperature and $\log{g}$ variations are almost in phase, while for the modes with $m>0$ there is a phase shift of typically $\pi$. For the modes with $l\geq3$ the variations became much too small to be detectable with this method (i.e. $\Delta T \leq 35 \mathrm{K}$). The variations for $l=1$,$m\pm1$ are shown in Figure 2b. While both modes have similar temperature amplitudes (m=+1: 539K, m=-1: 796K), the variations in surface gravity are drastically different: for $m=+1$ almost no variation (0.002$\mathrm{dex}$) is found, whereas for $m=-1$ $\log{g}$ varies as much as 0.043$\mathrm{dex}$.\\
The $T_{eff}$ amplitude predicted for the mode with $l=1, m=0$ matches the observed f1 pulsation (481.74s) quite well (s. Fig. 3).
 \begin{figure}[h]
\begin{center}
 \includegraphics [scale=.4] {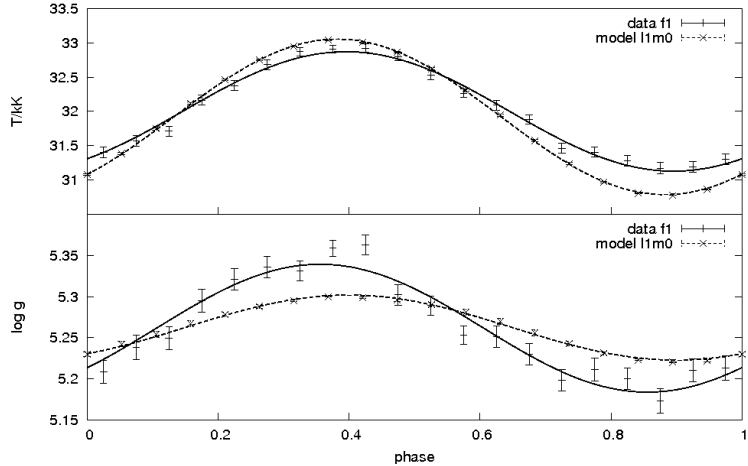}
\end{center}
\caption{\footnotesize{Temperature trend for the strongest mode determined from the Steward data and the adapted KYLIE model $l=1,m=0$ with error bars and sine fits are shown.}}
 \end{figure}
Although the surface gravity amplitude is too small we regard this mode as the most likely.\\
The second f2 ($P=475.61\mathrm{s}$) and third mode f3 ($P=475.76\mathrm{s}$) not only have almost the same period, but also the same amplitudes. In the modelling there were also only two modes $l=1,m=-1$ and $l=2,m=-1$ which produced variations in that range. Therefore we believe to know the quantum numbers of f2 and f3 but cannot distinguish between them.\\
Also for the fourth mode ($P=364.56\mathrm{s}$), there is actually only the mode with $l=2,m=-2$ left, which produces variations in the observed range.\\
The parameter range has to be further exploited to derive a consistent model.

\acknowledgements 
A.T. would like to thank the Royal Astronomic Society and the "Astronomische Gesellschaft" for their generous financial support.



\begin{thebibliography}{}
\bibitem{charpinet_1996}
Charpinet, S., Fontaine, G., Brassard D. et al., 1996, ApJ, 471, L103.
\bibitem{heber_1986}
Heber, U., 1986, A\&A, 155, 33.
\bibitem{heber_1999}
Heber, U., Reid, I.N. \& Werner K. et al., 1999, A\&A, 348, L25.
\bibitem{kawaler_1999}
Kawaler, S., 1999, in 11th. european Workshop on White Dwarfs, ASPC169, 158.
\bibitem{kilkenny_1997}
Kilkenny, D., Koen, C., O'Donoghue, D. et al., 1997, MNRAS, 285, 640.
\bibitem{kilkenny_1999}
Kilkenny, D., Koen, C., O'Donoghue, D. et al., 1999, MNRAS, 303, 525.
\bibitem{koen_1998}
Koen, C., O'Donoghue, D., Kilkenny, D. et al., 1998, MNRAS, 296, 317.
\bibitem{napiwotzki_1999}
Napiwotzki, R., 1999, A\&A, 350, 101.
\bibitem{otoole_2000}
O'Toole, S.J., Bedding, T.R., Kjeldsen, H. et al., 2000, ApJ, 537, L53.
\bibitem{otoole_2005}
O'Toole, S. J., Heber, U., Jeffery, C.S. et al., 2005, A\&A, 440, 667.
\bibitem{townsend_1997}
Townsend, R., 1997, PhD Thesis, University College London.
\end{thebibliography}
\end{document}